\documentclass[twocolumn]{revtex4}
\usepackage{amsmath,amssymb}
\usepackage{graphicx,psfrag,float}
\usepackage{mathrsfs,wasysym}

\DeclareSymbolFontAlphabet{\mathrsfs}{rsfs}
\DeclareMathAlphabet{\mathcal}{OMS}{cmsy}{m}{n}

\newcommand{\scri}{\mathrsfs{I}}

\newcommand{\be}{\begin{equation}}
\newcommand{\ee}{\end{equation}}

\def\tt{{\tilde{t}}}
\def\tr{{\tilde{r}}}
\def\tg{{\tilde{g}}}
\begin{document}

%%%%%%%%%%%%%%%%%%%%%%%%%%%%%%%%%%%%%%%%%%%%%%%%%%%%%%%%%%%%%%%%
%%%%%%%%%%%%%%%%%%%%%%%%%%%%%%%%%%%%%%%%%%%%%%%%%%%%%%%%%%%%%%%%

\title{A hyperboloidal study of tail decay rates for scalar and
  Yang-Mills fields} \author{An\i l Zengino\u{g}lu}
\email{anil@aei.mpg.de} \affiliation{Max-Planck-Institut f\"ur
  Gravitationsphysik, Albert-Einstein-Institut\\ Am M\"uhlenberg~1,
  D-14476 Golm, Germany}

\begin{abstract}
We investigate the asymptotic behavior of spherically symmetric
solutions to scalar wave and Yang-Mills equations on a Schwarzschild
background. The studies demonstrate the astrophysical relevance of
null infinity in predicting radiation signals for gravitational wave
detectors and show how test fields on unbounded domains in black hole
spacetimes can be simulated conveniently by numerically solving
hyperboloidal initial value problems.
\end{abstract}


\maketitle
\section{Introduction}
The purpose of this article is two-fold. First, we emphasize the
astrophysical relevance of null infinity in predicting gravitational
radiation signals as they are expected to be detected by gravitational
wave detectors. It has been demonstrated in \cite{Purrer:2004nq} that
a suitable feature to estimate the validity of the concept of null
infinity when predicting detector signals is the tail behavior. In
this article, we work out this idea further by discussing the observer
dependence of tails for solutions to scalar wave and Yang-Mills
equations on a Schwarzschild background.  Secondly, it is demonstrated
that test fields at null infinity can be conveniently studied
numerically by solving Cauchy problems once a suitable choice of gauge
for the background spacetime has been made.

The question about the astrophysical relevance of null infinity is a
conceptual one. Mathematically, it is only at null infinity that one
can give an unambiguous definition of gravitational radiation. Many
physicists, however, are uncomfortable with the notion of null
infinity because the detectors are at a finite distance away from the
source and move along timelike curves while idealized observers at
null infinity are infinitely far away from the source and move along
null curves. These issues are addressed within the article on concrete
examples.

Our aim is not to discuss the gauge dependency of radiation extraction
or the influence of finite distance boundary conditions on the
numerical solution as has been done for example in \cite{Pazos06,
  Rinne07}. We will rather numerically study a genuinely physical
phenomenon, namely the crucial dependence of a certain feature in the
radiation signal on the observers distance from the source
\cite{Winicour92,Purrer:2004nq}.

There is compelling evidence suggesting that, asymptotically with
respect to an appropriate time coordinate, the gravitational radiation
flux from a perturbed black hole decays polynomially in time
\cite{Ching95, Gundlach93a, Gundlach93b, Price72, Dafermos04b,
  Dafermos05}. This polynomial decay can be regarded as being due to
back-scatter of gravitational radiation off the curved background
spacetime.  The back-scattering effect is very small but it may, in
principle, be detected by future generation gravitational wave
detectors \cite{Blanchet95}. The decay rate of the tail may be of
particular interest in testing fundamental predictions of general
relativity because uniqueness theorems suggest that all features of
black holes other than mass, charge and angular momentum will decay
according to this rate independent of the details of
the collapse process. The tail plays an important role also in the
mathematical discussion of stability of black hole spacetimes and the
cosmic censorship conjecture \cite{Dafermos04b, Dafermos05}.

It turns out that the gravitational radiation flux of a general
perturbation has a faster decay on the event horizon and on timelike
surfaces than on null infinity \cite{Barack:1998bv, Barack:1998bw,
  Bonnor66, Price72, Burko:1997tb, Gundlach93a, Dafermos05}. This
difference raises the question of which decay is the relevant one with
respect to our gravitational wave detectors. P{\"u}rrer {\it et al} have
demonstrated in \cite{Purrer:2004nq} that the relevant decay rate in
the so-called astrophysical zone \cite{Leaver:1986gd} is the one
measured at null infinity. Their study is based on a characteristic
approach and discusses critical collapse for self-gravitating massless
scalar fields in spherical symmetry.

We are interested in a detailed study of the observer dependence of
decay rates including null infinity and assume the background to be
given. We compare a powerful mathematical estimate on decay rates for
scalar fields suggested in \cite{Szpak07a,Szpak07} with numerical
calculations and check decay rates for scalar fields with
non-vanishing angular momentum. We also study the tail behavior for
Yang-Mills fields where a breakdown of linear perturbation theory has
been demonstrated at future timelike infinity \cite{Bizon:2007xa}. Our
study indicates a similar breakdown of the linear analysis along
future null infinity.

Additionally, we show how test fields on black hole spacetimes can be
studied numerically by including null infinity while solving Cauchy
problems. Accurate studies of fields on black hole spacetimes
including null infinity have been restricted to the characteristic
approach until now \cite{Gundlach93a, Husa:2001pk,
  Campanelli:2000in}. The characteristic approach, however, has
certain difficulties mainly due to problems related to the rigidity in
the choice of the underlying gauge \cite{Winicour05}. The method
presented in this article is quite flexible and allows us to use
standard discretization techniques commonly applied in numerical
relativity. The feature that makes the accuracy of our calculations
possible is that the numerical outer boundary coincides with null
infinity so that, in the continuum problem, no boundary conditions are
required and no resolution loss appears in the physical part of the
conformal extension \cite{Andersson02a, Frauendiener98b, Husa05,
  Zeng07b}.
\section{The conformal method}
We want to study solutions to the scalar wave equation
\be \label{eq:scalar} \widetilde{\Box}\tilde{\Phi}=0,\ee for a scalar
field $\tilde{\Phi}$ in a Schwarzschild spacetime. To include null
infinity in the computational domain in a manifestly regular fashion,
we apply the conformal method introduced by Penrose \cite{Penrose63,
  Penrose64, Penrose65} (see also \cite{Frauendiener04, Husa02b,
  Friedrich02, Huebner96}). A conformal rescaling $g=\Omega^2\tg$ of
the physical metric $\tg$ with a conformal factor $\Omega>0$ implies
the transformation
\[ \left(\Box -\frac{1}{6} R \right) \Phi = \Omega^{-3} \left(\tilde{\Box} -
\frac{1}{6}\tilde{R}\right) \tilde{\Phi}, \quad \mathrm{with} \quad
\Phi = \frac{\tilde{\Phi}}{\Omega}, \] where $R$ and $\tilde{R}$ are
the Ricci scalars of the rescaled and the physical metrics $g$ and
$\tg$ respectively. The equation (\ref{eq:scalar}) with respect to the
rescaled metric then becomes \be\label{eq:solve} \Box\Phi -\frac{1}{6}
R \Phi = 0.\ee We are also interested in solutions to the Yang-Mills
equations, \be\label{eq:ym}D\ast F := d\ast F + A\wedge F = 0,\ee
where $F=d A + A\wedge A$ is the Yang-Mills curvature and $A$ is the
Yang-Mills connection. The Yang-Mills equations are conformally
invariant. We can therefore study the system (\ref{eq:ym}) directly in
a conformally rescaled spacetime.

Schwarzschild spacetime is weakly asymptotically simple, implying that
a suitably rescaled metric is smoothly extendable through null
infinity. We can solve the equations (\ref{eq:solve}) and
(\ref{eq:ym}) in a conformally extended Schwarzschild spacetime by
allowing the conformal factor to vanish in a certain way provided that
the initial data has sufficiently fast fall-off \cite{Geroch77}.

To solve an initial value problem for (\ref{eq:solve}) and
(\ref{eq:ym}) numerically on a Schwarzschild spacetime, we need to
choose a coordinate system that is suitable for numerical calculations
and covers that part of the Schwarzschild-Kruskal manifold we are
interested in, namely the domain extending from a neighborhood of the
event horizon up to future null infinity. Our coordinates should also
be adapted to a natural family of observers, each being thought as
representing gravitational wave detectors at constant distances away
from the black hole.

\begin{figure}[ht]
  \flushright 
  \psfrag{ip}{$i^+$} \psfrag{scrp}{$\scri^+$}
  \psfrag{hor}{$\mathcal{H}$} \psfrag{sing}{\scriptsize{singularity}}
  \includegraphics[height=0.21\textheight]{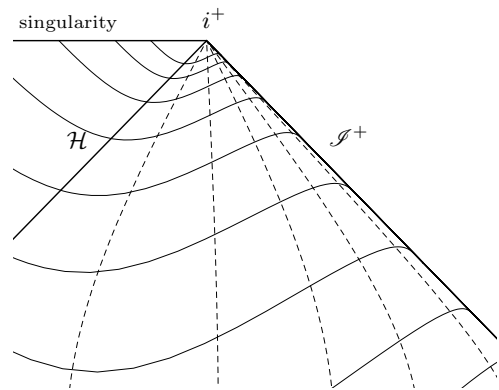}\hspace{1cm}
  \caption{Penrose diagram of a CMC-foliation in Schwarzschild
    spacetime with $m=1/2$, $C=1, \tilde{K}=2$. The dashed lines
    represent Killing observers. \label{fig:1}}
\end{figure}

A convenient foliation covering the extended Schwarzschild spacetime
up to future null infinity satisfying the above requirements can be
found in the class of spherically symmetric constant mean curvature
(CMC) foliations of Schwarzschild spacetime \cite{Brill80, Gentle00,
  MalecMurch03}. The transformation from the standard Schwarzschild
time coordinate to the time coordinate of a spherically symmetric
CMC-foliation can be written as \be\label{eq:time}
t=\tilde{t}-h(\tilde{r}),\ee where $\tilde{r}$ is the standard
Schwarzschild area radius, $h(\tilde{r})$ is the height function and
$\tt$ is the standard Schwarzschild time coordinate. The derivative of
the height function is given by \be\label{eq:height}
h'(\tr)=\frac{\frac{\tilde{K}\tilde{r}^3}{3} - C}
{\left(1-\frac{2m}{\tilde{r}}\right)\tilde{P}(\tilde{r})}, \ee with
\[ \widetilde{P}(\tr) :=
\sqrt{\left(\frac{\tilde{K}\tr^3}{3}-C\right)^2+
  \left(1-\frac{2m}{\tr}\right)\tr^4}.\] The mass of the Schwarzschild
black hole is denoted by $m$. The foliation parameters are the mean
extrinsic curvature $\tilde{K}$, and a constant of integration
$C$. The global behavior of CMC-surfaces depends on the foliation
parameters. We choose the parameters such that the surfaces of the
foliation come from future null infinity, pass the event horizon above
the bifurcation sphere and run into the future singularity as depicted
in Fig.~\ref{fig:1}.

The transformation (\ref{eq:time}) implies that the coordinates of the
foliation are adapted to time symmetry in the sense that the timelike
Killing vector field is given by $\partial_t$. To keep this property
for the conformally rescaled metric, we perform the conformal
compactification such that the conformal factor is time-independent
which we refer to as conformal fixing. The Schwarzschild metric in a
CMC-foliation and a conformal fixing gauge can be written as
\cite{Zeng07b}
\begin{eqnarray} \label{cmc_ss} g &=&
-\left(1-\frac{2 m\Omega}{r}\right)\Omega^2 dt^2 - 
\frac{2\left(\tilde{K}r^3/3-C\Omega^3\right)}{P(r)}\,dt dr + \nonumber
\\ &&+ \frac{r^4}{P^2(r)}\,dr^2 + r^2\,d\sigma^2, \end{eqnarray} 
where $d\sigma^2$ is the standard metric on the unit sphere,
\[ P(r):= \sqrt{\left(\frac{\tilde{K}r^3}{3} -
C\Omega^3\right)^2 + \left(1-\frac{2 m (1-r)}{r}\right)\Omega^2
r^4},\] and $\Omega=1-r$. Note that conformal fixing on a
hyperboloidal foliation implies scri-fixing, that is, the coordinate
location of null infinity in a conformal fixing gauge is independent
of the time coordinate. The compactifying coordinate $r$ is related to
the standard Schwarzschild area-radius $\tr$ via \be\label{eq:compr}
\tr = \frac{r}{1-r} = \frac{r}{\Omega}.\ee 

The timelike curves with constant spatial coordinates can be regarded
as worldlines of natural observers at constant
distances away from the source. As can be seen from Fig.~\ref{fig:1}
and from $g(\partial_t,\partial_t)|_{\scri^+}=0$, the worldline of such
observers becomes null at null infinity.
\section{The initial value problem}
\subsection{The continuum problem}
We write the metric (\ref{cmc_ss}) as \[ g = \left(-\alpha^2 +
\gamma^2 \beta^2\right)\,dt^2 + 2 \gamma^2\beta\, dt\,dr + \gamma^2
\,dr^2 + r^2 \,d\sigma^2. \] The lapse $\alpha$, the shift $\beta$,
and the spatial metric function $\gamma$ read
\be \label{eq:cmc_ss_3+1} \alpha=\frac{P(r)}{r^2},\quad \beta =
\left(-\frac{\tilde{K} r}{3}+\frac{C\,\Omega^3}{r^2}\right)\,\alpha,
\quad \gamma = \frac{1}{\alpha}.\ee We apply the method of separation
of variables to (\ref{eq:solve}). We write
$\Phi(t,r,\vartheta,\varphi) = \phi(t,r) Y_{lm}(\vartheta,\varphi)$
with $Y_{lm}$ being the usual spherical harmonics and define the
auxiliary variables
\be\label{eq:aux} \psi:= \partial_r \phi, \quad\mathrm{and}\quad \pi :=
\frac{\gamma}{\alpha}(\partial_t\phi-\beta\partial_r \phi). \ee Then
the following linear, symmetric hyperbolic system of evolution
equations for $\phi(t,r)$ is obtained 
\begin{eqnarray}\label{eq:wave}
\partial_t \phi &=&\frac{\alpha}{\gamma}\,\pi + \beta\psi, \nonumber \\
\partial_t \psi &=&\partial_r\left(\frac{\alpha}{\gamma}\pi +
\beta\psi\right), \\
\partial_t \pi &=&
\frac{1}{r^2}\,\partial_r\left(r^2\left(\frac{\alpha}{\gamma}
\psi+\beta\pi\right)\right) - \alpha\gamma\left( \frac{1}{6}\,R \phi +
\frac{l(l+1)}{r^2}\phi\right) \nonumber. 
\end{eqnarray}
The Ricci scalar of the metric (\ref{cmc_ss}) reads
\[ R=\frac{12 \Omega}{r^2} \left(r+m(2r-1)\right).\]
For the Yang-Mills equations (\ref{eq:ym}) we consider the gauge group
$SU(2)$ and make the following spherically symmetric ansatz for the
connection \cite{Bartnik92, Choptuik96}
\[ A = (\phi+1) \,\sigma_1 \,d\vartheta + (\cos\vartheta \,\sigma_3+(\phi+1)
\sin\vartheta \,\sigma_2)\, d\varphi,\] where the $\sigma_i$ are Pauli
matrices and $\phi=\phi(t,r)$. With the above choice, a vacuum state
for the Yang-Mills field is given by $\phi=0$. The auxiliary variables
are defined as in (\ref{eq:aux}). The only difference to the system
(\ref{eq:wave}) is the equation for the time derivative of $\pi$ which
becomes \be\label{eq:ym_pi} \partial_t \pi =
\partial_r\left(\frac{\alpha}{\gamma} \psi+\beta\pi\right) -
\frac{\alpha\gamma}{r^2}\,\phi(1+\phi)(2+\phi).\ee We choose an
approximately outgoing, compactly supported Gaussian pulse for initial
data. On the initial hypersurface such data satisfies
\[ \partial_t (r\phi) + c_+ \partial_r(r\phi) = 0, \]
where $c_+$ is the outgoing characteristic speed given by $c_+ =
-\beta + \alpha/\gamma$. The data reads for
$r<r_{\mathrm{out}}=\mathrm{const.}<1$  as follows
\begin{eqnarray} \label{eq:id}
\phi(0,r) &=& a\,e^{-(r-r_c)^2/\sigma^2}, \nonumber \\ \psi(0,r) &=&
-\frac{2\,(r-r_c)}{\sigma^2}\,\phi(0,r), \nonumber\\ \pi (0,r) &=&
-\psi(0,r) - \frac{\phi(0,r)}{r}
\left(1-\frac{\beta\gamma}{\alpha}\right),
\end{eqnarray}
where $r_c$ is the center of the Gaussian pulse, $a$ is its amplitude
and $\sigma$ is its width. The data is set to zero for $r\geq
r_{\mathrm{out}}$. 

The initial hypersurface is hyperboloidal, meaning that it is a smooth
spacelike hypersurface extending through null infinity. The problem
consisting of initial data given on a hyperboloidal surface and
evolution equations is called a hyperboloidal initial value problem
\cite{Friedrich83a}. Note that no boundary conditions are needed in
the continuum problem.
\subsection{The numerical method}
We may solve hyperboloidal initial value problems for scalar
wave and Yang-Mills equations using standard numerical
techniques. I used the method of lines with fourth order Runge-Kutta time
integration and finite differencing with sixth order accurate
stencils. The inner boundary is chosen to be a spacelike surface
inside the event horizon so that one can use an excision
technique. The outer boundary is at null infinity where one-sided
finite differencing is applied. The lack of outer boundary conditions
makes the numerical implementation simpler than in the case where an
artificial timelike outer boundary is introduced.

\begin{figure}[ht]
  \center
  \psfrag{t}{$t$}  \psfrag{Q}{$Q$}
  \includegraphics[height=0.17\textheight]{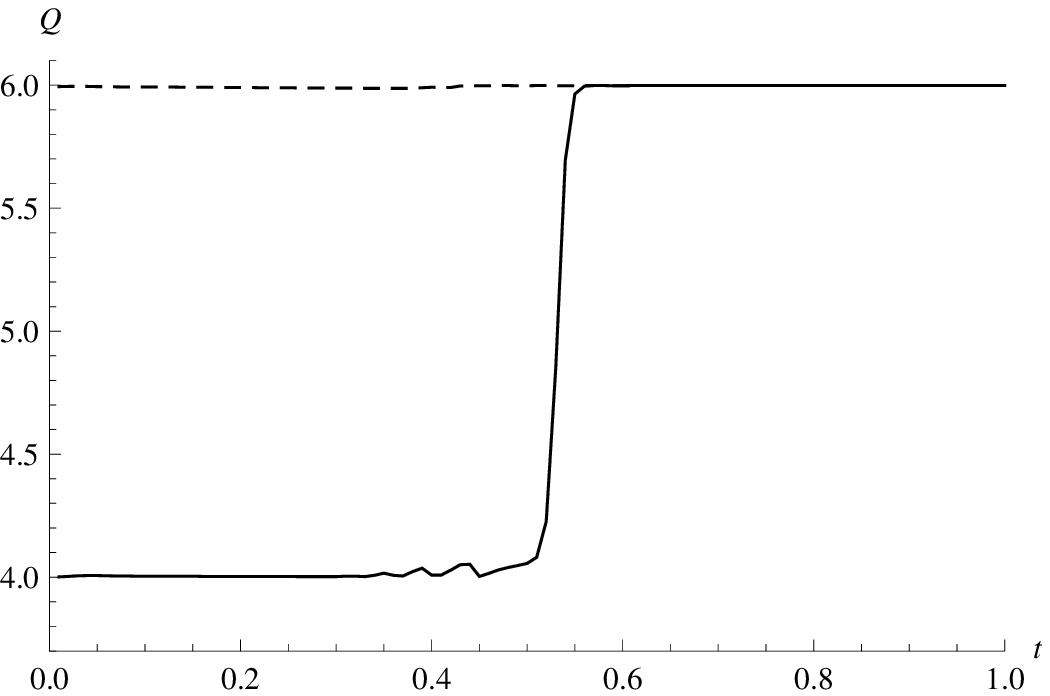}\hspace{1cm}
  \caption{Convergence in the L2-norm for numerical solutions to
    (\ref{eq:wave}). The convergence factor $Q$ is calculated by
    \mbox{$Q=\log_2\frac{ \| \phi^{low}-\phi^{med}\|}{\|
        \phi^{med}-\phi^{high}\|} $}.\label{fig:2}}
\end{figure}

The convergence of the code can be seen in Fig.~\ref{fig:2}. For this
plot, a three level convergence analysis has been performed with 1010,
2020 and 4040 grid cells. The Courant factor $\triangle t/\triangle
r$ is chosen to be $0.5$. The convergence in the L2-norm has been
plotted for a solution to the hyperboloidal initial value problem for
the scalar wave equation. The corresponding plot for the Yang-Mills
system is similar. We see that the code switches from fourth to sixth
order convergence after a short time. This seems to be due to the
initially dominating error from the fourth order time integration, and
indeed, choosing a Courant-factor 10 times smaller results in the
dashed curve which indicates sixth order convergence from the
start. As we are interested in long time evolutions, the larger
Courant factor is chosen. The convergence factors in the maximum and
the L1-norms behave similarly.

For the results presented below the number of grid cells is chosen to
be 10020. The spatial simulation domain in grid coordinates is given by
$r\in[0.499,1]$. The Courant factor is 0.2. The parameters of the
spacetime and the foliation are $m=0.5,\tilde{K}=0.5,C=1$.
\section{Results}
\subsection{Tails for the scalar wave equation}
The late-time decay of a solution to the scalar wave equation is
expected to have the form
\[ \lim_{t\to\infty}\phi(r,t) = C t^{p}, \]
where $C$ depends on $r$ only and $p$ is the decay rate with
$p<0$. For a solution with vanishing angular momentum, that is, $l=0$,
we expect along the event horizon and along timelike surfaces a decay
rate $p=-3$ \cite{Price72}. At null infinity, however, the decay rate
is expected to be $p=-2$ \cite{Bonnor66,Gundlach93a}. We would like to
know which decay rate is the relevant one in the context of
astrophysical predictions and have a quantitative understanding of the
dependence of the decay rate on the observers location
\cite{Purrer:2004nq}.

\subsubsection{Comparison between mathematical and numerical results}
Recently, an optimal pointwise decay estimate has been suggested, but
not yet proven, for small spherically symmetric solutions to nonlinear
wave equations with a potential \cite{Szpak07a, Szpak07}. Applying
this estimate to our linear problem gives \be\label{eq:pointwise}
|\tilde{\phi}| \leq
\frac{C_1}{(C_2+\tt+\tr_\ast)(C_3+\tt-\tr_\ast)^2},\ee where
$(C_1,C_2,C_3)$ are constants, $\tr_\ast:=\tr+2 m \log (\tr-2m)$ and
$\tt$ is the standard Schwarzschild time coordinate. This estimate
captures the asymptotic behavior of the solution near null infinity
as well as at finite distances from the black hole. The corresponding
decay rates can easily be read off from (\ref{eq:pointwise}). We want
to compare the mathematical estimate with the numerical calculation
with respect to the prediction of the decay rate. We calculate the
function \be\label{eq:exponent} p(t,\tr) = \frac{d\ln |\phi(t,\tr)|}{d\ln
  t}, \ee which becomes asymptotically the exponent of the polynomial
decay of the solution.  For the comparison, we need to write down the
mathematical estimate (\ref{eq:pointwise}) in the coordinates of our
numerical calculation. The coordinate transformation is given by
(\ref{eq:time}). The height function in this transformation is not
given in explicit form, but as we are mainly interested in the
asymptotic behavior of the solution in the far-field zone we can use
a Taylor expansion of (\ref{eq:height}) as $\tr\to\infty$. We get
\[ h(\tr) = \tr_\ast + \frac{1}{\tr}\left(\frac{9}{2K^2}-4m^2\right) -
\frac{4m^3}{\tr^2} + O\left(\frac{1}{\tr^3}\right).\] 
\begin{figure}[ht]
  \center
  \psfrag{t}{$t/m$}  \psfrag{p}{$p(t,R)$}
  \includegraphics[height=0.21\textheight]{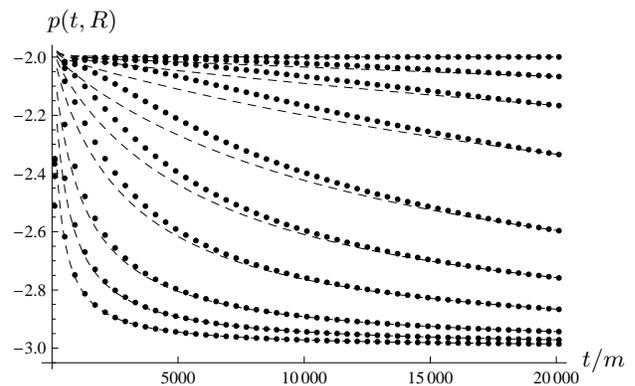}\hspace{1cm}
  \caption{Comparison between the mathematical pointwise estimate
    (\ref{eq:pointwise}) and a numerical calculation. Each curve
    depicts the function $p(t,R)$ calculated as in (\ref{eq:exponent})
    for the following values \mbox{$R/m=(\infty,40000,20000,10000,4000,
    2000,1000,400,200,100)$}. \label{fig:3}}
\end{figure}

The comparison has been plotted in Fig.~\ref{fig:3}. Each curve in the
figure gives the function $p(t,R)$ for a constant value of $R$. The
dashed lines represent the mathematical estimate which have been
fitted at late times to some representative data points from the
numerical simulation. We set $C_1=C_2=1$ and treat $C_3$ as a fitting
parameter along each curve. Considering that the formula
(\ref{eq:pointwise}) is only an estimate and not an approximation to
the solution, we can say that it is remarkable how well the
qualitative behavior of the asymptotic decay is captured by it.
\subsubsection{Astrophysical relevance of null infinity for the decay rate}
Now we can discuss the question which decay rate is the relevant one
for gravitational wave detectors. We think of each curve in
Fig.~\ref{fig:3} as representing the measurement made by a detector
at some constant distance away from the source. In a Schwarzschild
spacetime we have a length scale at our disposal given by $m$, the
mass of the black hole, so that we can gain an intuition for the
scales by considering the distances of some typical astronomical
objects in terms of geometric units. The closest candidate for a
supermassive black hole is Sgr $A^\ast$ at the center of our
galaxy. Its mass is about $3.7\times 10^6 M_{\astrosun}$ and it is
about $26000$ light years away. In geometric units, this corresponds
to roughly $10^{10}m$. Another close example is the first serious
candidate for a stellar size black hole, namely the compact object in
the binary system Cygnus X-1. The total mass of the binary system is
about $50M_{\astrosun}$ and its distance to the Earth is about $8000$
light years. This gives in geometric units a distance of roughly
$10^{15}m$.  Now consider in Fig.~\ref{fig:3} the curve that is
closest to what has been calculated along $\scri^+$. Its distance to
the black hole in geometric units is $4\times 10^{4} m$. While
this is quite far for numerical calculations, it is obviously still
very close compared to astronomical distances. Even for such
comparably small distances we see that the relevant decay rate is the
one measured at null infinity.

One can of course argue that the decay rate along any timelike surface
at a finite distance will eventually approach $-3$ if one waits long
enough. It is clear from Fig.~\ref{fig:3} and from the estimate
(\ref{eq:pointwise}) that the time scale for this to happen depends on
the distance of the source. If we have a source that is, for example,
1000 light years away from the Earth, then the decay rate will come
close to $-3$ only after about $10000$ years. This is in accordance
with the argument that the mathematical idealization of the region
$\tr\gg \tt$ for radiative properties of fields corresponds to null
infinity while for $\tt\gg \tr_\ast$ the correct idealization is
timelike infinity \cite{Frauendiener98c, Leaver:1986gd,
  Purrer:2004nq}.  Fig.~\ref{fig:3} depicts the transition zone
between these two regions. We should mention that it is
extremely unlikely that the polynomially weakening tail signal can be
followed such a long time that the decay rate at timelike infinity
will be relevant. It is rather likely that if ever the tail decay rate
is measured by direct observation, the measurement will be possible
only for a short time.

The above discussion recapitulates that $\scri^+$ is the appropriate
idealization for an observer's location in an asymptotically flat
spacetime. One should, however, be aware that future generation
detectors will observe variations in the radiation signal due to
cosmological effects when the sources are considerably far away in
comparison to the size of the known universe. In this case, one might
reconsider the relationship between the different idealizations of
isolated systems and cosmological spacetimes.
\subsubsection{Construction of a physically motivated time coordinate}
In our discussions and conclusions we use the coordinate time $t$. In
this subsection we want to briefly elaborate that the same conclusions
can be drawn with respect a physically motivated time coordinate.

\begin{figure}[ht]
  \flushright 
  \psfrag{t}{$t/m$}  \psfrag{r}{$\tr/m$}
  \includegraphics[height=0.21\textheight]{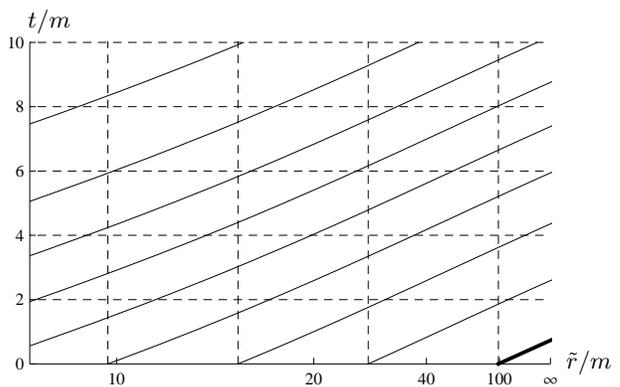}\hspace{1cm}
  \caption{Grid diagram depicting outgoing null curves on the grid. To
    synchronize observers from $100m$ to infinity, we need to take
    care of the shift in time given by the height of the thick
    curve. The diagram shows that the required shift is negligible in
    comparison to the time scales we are interested in.
\label{fig:4}}
\end{figure}

We have chosen our foliation suitably such that worldlines of natural
observers of gravitational radiation are represented by timelike
coordinate lines. The parametrization of these curves by $t$, however,
can be regarded as being arbitrary. There are two steps to be taken in
the construction of a more physically motivated time coordinate. First,
we can synchronize our observers by an outgoing light pulse so that
they have a common starting time for their measurements.  This
starting time $t_R$ for an observer located at $\tr=R$ can be
calculated by
\[ \int_{t_R}^T dt = T-t_R=\int_{R}^{\infty} c_+(\tr)\,d\tr, \]
where $T$ is the time at infinity and $c_+$ is the outgoing speed of
light rays. It does not matter whether $c_+$ is calculated with
respect to the physical or the conformally rescaled metric because the
null cone structure is invariant under conformal rescalings. We depict
in Fig.~\ref{fig:4} null curves on our grid. We see that the
difference in the initial time shift required for the synchronization
is almost negligible for observers sufficiently far away. The reason
for this is the fact that surfaces of a future hyperboloidal foliation
follow closely outgoing null surfaces. Note that in a Cauchy-type
foliation, the synchronizing shift in time would be calculated with
respect to a far away observer at a finite distance. The required
shift would then be on the order of the distance of that observer.

The next step for the construction of a physically motivated time
parameter would be the reparametrization of the time coordinate so
that the measurement by a given observer is plotted in proper time.
The relation of the proper time of a Killing observer at $\tr=R$ to
the coordinate time is given by
\[ \triangle s=\int \sqrt{-\tg_{tt}} dt =
\left(1-\frac{2m}{R}\right)^{1/2} \triangle t.\] We see that at
infinity proper time corresponds to coordinate time. The rescaling
makes only a small difference for observers far away.

A transformation of the results presented in this article according to
the choice of such a physically motivated time coordinate does not
change the conclusions for far away observers. Therefore, and because
of simplicity, the plots will be given in the local time coordinate of
the numerical solution. Note, however, that we do not plot the decay
rates for observers near the horizon in our diagrams although our
coordinates do allow this as well.
\subsubsection{A remark on finite distance extractions}
In numerical calculations where an artificial timelike outer boundary
is introduced, one extracts gravitational radiation information from
the numerical solution at finite distances away from the source. When
the extracted information does not vary in an unexpected manner with
extraction radius or algorithm, one suggests that the calculation
delivers a reliable answer.  This practice can be regarded as safe as
long as one has a theoretical understanding of the expected radiation
signal which delivers tests of the result, such as the peeling
behavior, conservation of mass or waveforms from approximation
methods to compare with \cite{Boyle:2007ft, Hannam:2007ik}. A naive
interpretation of finite distance extractions without theoretical
knowledge, however, can be misleading in terms of results as well as
error estimates.
\begin{figure}[ht]
  \center
  \psfrag{p}{$p(T,\tr)$}\psfrag{r}{$\tr/m$}
  \includegraphics[height=0.21\textheight]{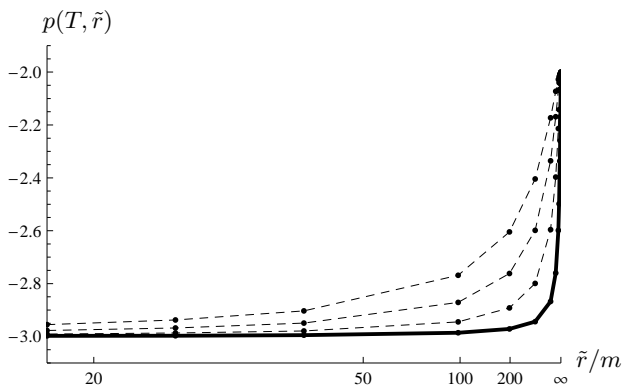}\hspace{1cm}
  \caption{Observer dependence of the tail decay rates at subsequent
    times. The thick curve corresponds to $T=20000m$. \label{fig:5}}
\end{figure}

The decay rates at late times provide an example for a case which
demands caution. In Fig.~\ref{fig:5} we present a different way
of looking on the same data as given in Fig.~\ref{fig:3}. Here,
decay rates at different finite distances have been plotted on
subsequent time steps. The thick curve corresponds to the observer
dependence of the decay rates at late times. We see that a direct
Richardson extrapolation based on finite distance decay rates would
not only deliver a wrong result, but also a wrong error estimate.
\subsubsection{Scalar field with angular momentum}
\begin{figure}[ht]
  \center
  \psfrag{t}{$t/m$}  \psfrag{p}{$p(t,R)$}
  \includegraphics[height=0.21\textheight]{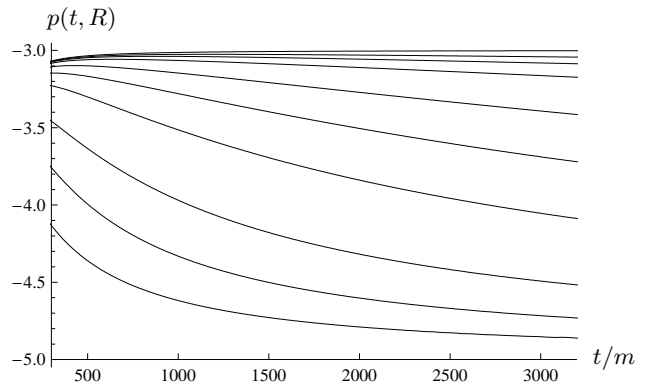}\hspace{1cm}
  \caption{Decay rates for a solution to the scalar wave equation on a
    Schwarzschild background with $l=1$ plotted for the same observers
    as for Fig.~\ref{fig:3}. This plot covers a shorter time than
    Fig.~\ref{fig:3} due to the faster fall-off of the
    field. \label{fig:6}}
\end{figure}

The difference in the expected tail behavior at null infinity and at
finite distances becomes larger with $l$. At finite distances one
expects the exponent to be \mbox{$p = -2l-3$}, whereas at null
infinity $p=-l-2$. In Fig.~\ref{fig:6} we plot the decay rates for
$l=1$ after the quasinormal mode ringing is over and the tail part of
the solution remains. In this case, there is no conjecture of an
optimal pointwise decay estimate yet available in the literature, so
we only plot the result of the numerical calculation.

The qualitative behavior depicted in Fig.~\ref{fig:6} is similar to
the one depicted in Fig.~\ref{fig:3} which suggests that an
estimate similar to (\ref{eq:pointwise}) with modified exponents can
be expected also for higher multipoles.
\subsection{Tails for the Yang-Mills equation}
The conformal method can be applied to study other fields as well. The
basic requirements are that the equations transform in a well-defined
manner under conformal rescalings of the metric and the fields have
a suitable asymptotic behavior. Yang-Mills equations are conformally
invariant so that we can study them directly with the conformal
method.

The tail behavior for Yang-Mills equations have been studied in
\cite{Cai99} using linear perturbation analysis. They predict $p=-5$
at future timelike infinity and $p=-3$ at future null infinity. Bizon
et.~al.~showed in \cite{Bizon:2007xa} that the linear perturbation
theory applied in \cite{Cai99} does not capture the correct tail
behavior for Yang-Mills equations where nonlinear terms appear
(\ref{eq:ym_pi}). Instead they argued and numerically demonstrated
that the tail exponent at timelike infinity is $p=-4$ both for
Minkowski and for Schwarzschild spacetimes.
\begin{figure}[ht]
  \center
  \psfrag{t}{$t/m$}  \psfrag{p}{$p(t,R)$}
  \includegraphics[height=0.21\textheight]{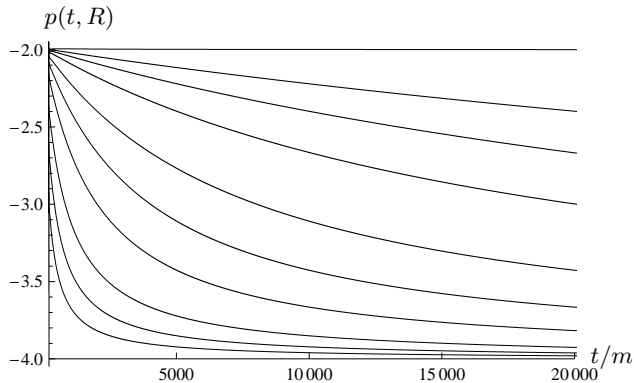}\hspace{1cm}
  \caption{Decay rates for a solution to Yang-Mills equations on a
    Schwarzschild background. Each curve depicts the function $p(t,R)$
    calculated as in (\ref{eq:exponent}) for the following values
    \mbox{$R/m=(\infty,40000,20000,10000,4000,2000,1000,400,200,100)$}.
     \label{fig:7}}
\end{figure}

We plot in Fig.~\ref{fig:7} the observer dependence of decay rates
for a solution of the Yang-Mills equations. Near future timelike
infinity a decay rate of \mbox{$p=-4$} is observed in accordance with
\cite{Bizon:2007xa}. One might expect that the prediction of
\cite{Cai99} at future null infinity will also be wrong by the same
rate which indeed seems to be the case. As seen in Fig.~\ref{fig:7},
the numerical calculation suggests a decay rate of $p=-2$ for
Yang-Mills fields at null infinity on a Schwarzschild background.
\section{Conclusions}
We investigated the asymptotic behavior of spherically symmetric
solutions to scalar wave and Yang-Mills equations on a Schwarzschild
background. The investigation develops an idea presented in
\cite{Purrer:2004nq} where, among others, decay rates have been used
to estimate the validity of the concept of null infinity for
predicting detector signals. Our investigation focuses on the observer
dependence of decay rates. In the case of scalar fields with vanishing
angular momentum, a powerful mathematical estimate presented in
\cite{Szpak07a,Szpak07} could be compared with a numerical
solution. For scalar fields with angular momentum, no such estimate is
yet available, but the numerical results suggest that a similar
estimate with modified exponents should be valid. For Yang-Mills
fields the decay rates near timelike infinity and null infinity have
been studied. The breakdown of linear perturbation theory for
predicting decay rates near timelike infinity in case of Yang-Mills
fields as discovered in \cite{Bizon:2007xa} could be numerically
supported also for decay rates along null infinity. 

On the example of the above studies, we emphasized the astrophysical
relevance of null infinity. The examples are intended to contribute to
the clarification of the question of to what extent null infinity can
be regarded as a good idealization for far away observers. The studies
also deliver a cautious remark against naive interpretations of finite
distance extractions. One needs to be careful in wave extraction not
only due to gauge issues in extraction methods and influences from the
outer boundary, but also due to the fact that the extraction radius in
standard numerical calculations is very small in comparison with
astronomical distances. While the tail is certainly a special feature
in that respect, there might be other features in the radiation signal
that show a non-trivial dependence on the extraction surface. Advanced
numerical studies including null infinity should be carried out to
clarify this issue as has been initiated within the characteristic
approach \cite{Winicour92}.

The numerical studies of null infinity presented in this article have
been performed by solving hyperboloidal initial value problems
\cite{Geroch77, Friedrich83a}. This method seems to be quite powerful
to study radiative properties of test fields at null infinity once a
suitable class of gauges has been chosen \cite{Andersson02a,
  Frauendiener98b, Husa05}. It also promises to be more general than
the characteristic approach due to the large freedom involved in the
choice of hyperboloidal gauges. In comparison to the standard initial
boundary value problem, the numerical implementation is simpler
because one does not need to mathematically construct or numerically
implement boundary conditions and data. In addition, the conclusions
that can be drawn from the numerical simulation are more powerful as
radiative properties of fields can be studied all the way up to null
infinity.

In future work, the hyperboloidal initial value problem should be
further studied, for example, for different test fields or on
different backgrounds such as on Kerr spacetime using a conformal
fixing gauge as presented in \cite{Zeng07b}. The most important
challenge in numerical studies of hyperboloidal initial value problems
is to devise a successful numerical implementation of the conformally
compactified Einstein equations including null infinity. While there
are some recent ideas on how to treat a hyperboloidal initial value
problem for the Einstein equations directly \cite{Andersson02a,
  Husa05, Moncrief06, Zeng-phd}, they have not yet been implemented
numerically for interesting cases. Studies of test fields might give
further insight into this challenging problem.
\begin{acknowledgments}
I thank Sascha Husa for his help with the calculations on various
stages of this research as well as for comments on the manuscript. I
would also like to thank Helmut Friedrich, Luciano Rezzolla and
Nikodem Szpak for discussions and support.
\end{acknowledgments}

\bibliography{references}\bibliographystyle{apsrmp}
\end{document}